\documentclass[12pt]{iopart}

\usepackage{graphicx}
\usepackage{amssymb}
\usepackage{color}

\newcommand{\ket}[1]{\ensuremath{|#1\rangle}}
\newcommand{\bra}[1]{\ensuremath{\langle#1|}}
\newcommand{\braket}[2]{\langle#1|#2\rangle}

\begin{document}

\title{Sequential measurement of displacement and conduction currents in electronic devices}

\author{Guillermo Albareda}
\address{Institut de Qu\'imica Te\`orica i Computacional and Departament de Qu\'imica F\'isica, Universitat de Barcelona, Barcelona, Spain}
\ead{albareda@ub.edu}
\author{Fabio Lorenzo Traversa}
\address{Department of Physics, University of California, San Diego, La Jolla, CA 92093-0319, USA}
\ead{ftraversa@physics.ucsd.edu}
\author{Abdelilah Benali}
\address{Department of Electronic Engineering, Autonomus University of Barcelona Bellaterra, Barcelona 08193, Spain}
\ead{abdelilah.bnl@gmail.com}
\vspace{10pt}
\begin{indented}
\item[]October 2015
\end{indented}

\begin{abstract}
The extension of the Ramo-Schockley-Pellegrini theorem for quantum systems allows to define a positive-operator valued measure (POVM) for the total conduction plus displacement electrical current.
The resulting current operator is characterized by two parameters, viz. the width of the associated Gaussian functions and the lapse of time between consecutive measurements. 
For large Gaussian dispersions and small time intervals, the operator obeys to a continuous weak-measurement scheme. 
Contrarily, in the limit of very narrow Gaussian widths and a single-shot measurement, the operator corresponds to a standard von Neumann (projective) measurement.
We have implemented the resulting measurement protocol into a quantum electron transport simulator.
Numerical results for a resonant tunneling diode show the great sensibility of current-voltage characteristics to different parameter configurations of the total current operator.
\end{abstract}


\vspace{2pc}


 

\section{Introduction}
At high frequencies, the time-dependent electrical current measured by an ammeter is made of two different contributions: the conduction and the displacement current~\cite{Jakson,MJavidMcGrawHillBook1963}.
Time scales experimentally accessible in electronic devices have been, up to now, much larger than the temporal widths of the pulses generated by conduction electrons, mainly because of parasitic RC elements~\cite{Taur}. 
Therefore, the computation of the displacement current has been mostly disregarded in the recent literature.
Nowadays, however, the growing demand for larger bandwidth in communication systems and new sensor applications will require devices to extend the operating frequency deep into the terahertz regime.
In this respect, many emerging devices are envisioned or have been already proved to operate close to or at a few terahertzs~\cite{Graphene, Vertical, Asada, Orihashi, Burke, Rangel}.
In this regime, the contribution of the displacement current to the total electrical current cannot be disregarded anymore, and hence a theoretical approach to describe this 
type of currents is necessary to evaluate the performance of novel electronic devices.

The seminal papers by Shockley~\cite {WShockleyJAP1938} and Ramo~\cite{SRamoPIRE1938} on vacuum tubes showed how the total (conduction plus displacement) 
current on a given surface can be better computed and understood through mathematical expressions involving a spatial integral over an arbitrary volume containing that surface. 
The works of Ramo and Shockley have been extended, to a greater or lesser extent, for solid state electronic devices~\cite{GCavalleri_NIM_1971,TWessegBerg_PS_1978,HunsukKimetal_SSE_1991,DYoderJAP_1996}. 
However, the existence of a very general version of the original works has not been proved until Pellegrini's contribution~\cite{BpellegriniPRB1986}. 
In what he called the electrokinematic theorem, Pellegrini provided a set of general expressions relating the (spatial and frequency dependent) dielectric constant, the conduction current, the (scalar and vector) 
electromagnetic potentials and an arbitrary irrotational field for a given volume. 
In order to acknowledge the relevant work of Pellegrini, hereafter we will use the name Ramo-Shockley-Pellegrini (RSP) theorem when referring to the computation of the total current using volume integrals.

Due to the successful application of the RSP theorem in classical scenarios (without frequency limitations), it seems natural to extend the applicability of the RSP theorems to quantum scenarios.  
In fact, the RSP theorem has been extended for quantum systems by Pellegrini himself~\cite{PellegriniIlnuevocimento1} and also by the authors of the present work~\cite{QuantumRSP} using, respectively, exact wavefunction and trajectory-based schemes. 
An important consideration was however obviated in these previous works: the effects of the measuring apparatus on the evolution of the quantum system. 
When the electrical current is being continuously monitored, particle dynamics is unavoidably affected through its interaction with the ammeter. This is known as the measurement backaction, and it is at the very heart of quantum mechanics.
In particular, for von Neumann measurement schemes~\cite{vonNeuman}, the effects of measurements can even lead to the suppression of dynamics~(see for example~\cite{zeno}).
Therefore, if one is aiming at provide a theoretical approach to quantify the \emph{measured} electrical current at high frequencies, and hence be able to make a fair comparison with experiments, 
it is mandatory to extend the previous works to account for the interaction of the ammeter with the system of interest.

Different approximations have been developed to compute AC currents in quantum systems and account at the same time for the ``collapse'' of the wavefunction~\cite{Butiker}. 
Most of these approaches use projective operators to describe the current measurement, assuming an ``instantaneous'' and uniform change on the potential and charges inside the simulation region. 
They are, though, limited to frequencies below hundreds of GHz~\cite{blanter2000PR}, implying intervals between consecutive measurements larger than the characteristic electron transit times. 
In order to overcome the GHz limit, novel approaches based on the ability of positive-operator valued measure (POVM) to provide information without significantly distorting the wavefunction are being developed to 
describe continuous (i.e., above hundreds of GHz) measurements~\cite {povms}. 
POVMs are specially suited when some aspect of a system is continually monitored (i.e. interacting with the environment) and have become increasingly important in the last decade, due mainly to the growing interest in the application 
of feedback control in quantum systems~\cite{feedback1,feedback2,feedback3}.

Starting from the RSP theorem for quantum systems, in this work we propose a theoretical weak measurement protocol to measure the total (conduction plus displacement) current in two-terminal electronic devices. 
This measurement protocol in combination with a time-dependent electron transport simulator allows the computation of the total (displacement and conduction) current in nanoscale electronic devices taking into account system-environment 
interactions.

After this introduction, in section~\ref{section_2}, we introduce the RSP theorem for quantum systems and derive the weak measurement protocol for the total current.
In section~\ref{section_3}, we present numerical results for a resonant tunneling diode.   
We conclude in section~\ref{section_4}.

\section{Modeling the ammeter through the Ramo-Shockley-Pellegrini theorem}
\label{section_2}
Before addressing the modeling of the ammeter-system interaction and deriving a weak-measure protocol for the total electrical current, let us here briefly describe the system of interest.
Consider an ensemble of interacting (spinless) particles under the action of an external electrical field and defined through the many-body state $\psi(\mathbf{r},t)$, 
where $\mathbf{r} = (\mathbf{r}_1,..,\mathbf{r}_{N})$ collectively denotes the position of $N$ particles (including electrons and ions) in the $3N$-dimensional configuration space. 
We assume that the state $\psi(\mathbf{r},t)$ describes an electronic device and that it is effectively governed by the following time-dependent Schr\"odinger equation:
\begin{equation}
i\frac{\partial}{\partial t}\psi(\mathbf{r},t) = H(\mathbf{r},t)   \psi(\mathbf{r},t) = \left(  K +  U(\mathbf{r},t) \right)  \psi(\mathbf{r},t),
\label{schrodinger}%
\end{equation}
where $K = \sum_k^N \frac{1}{2m_k} {\mathbf{p}}_k \cdot {\mathbf{p}}_k$ and ${\mathbf{p}}_k = -i\nabla_k$ are respectively the many-body kinetic energy and the $k$-th linear momentum operators written in the position representation. 
The term ${U}(\mathbf{r},t) = {U}_{cou}(\mathbf{r}) + \sum_{k=1}^{N} {U}_{ext}(\mathbf{r}_k,t)$ represents the (scalar) potential energy accounting for the Coulomb interaction $U_{cou}$ among the $N$ particles and also for
their interaction with an external electric field through $U_{ext}$. 
Throughout this work we use atomic units ($m_{e}=1$, $q_e=-1$, $\hbar =1$).

\subsection{The Ramo-Shockley-Pellegrini theorem for quantum systems}
\label{section_2_1}
We now address the question of what is the total (conduction plus displacement) current generated by the state $\psi(\mathbf{r},t)$ and obeying the effective Hamiltonian $H$ in Eq.~\eref{schrodinger}.
To this end, here we summarize the main result of Refs.~\cite{PellegriniIlnuevocimento1,QuantumRSP}.

\begin{figure}[th]
\includegraphics[width=\textwidth]{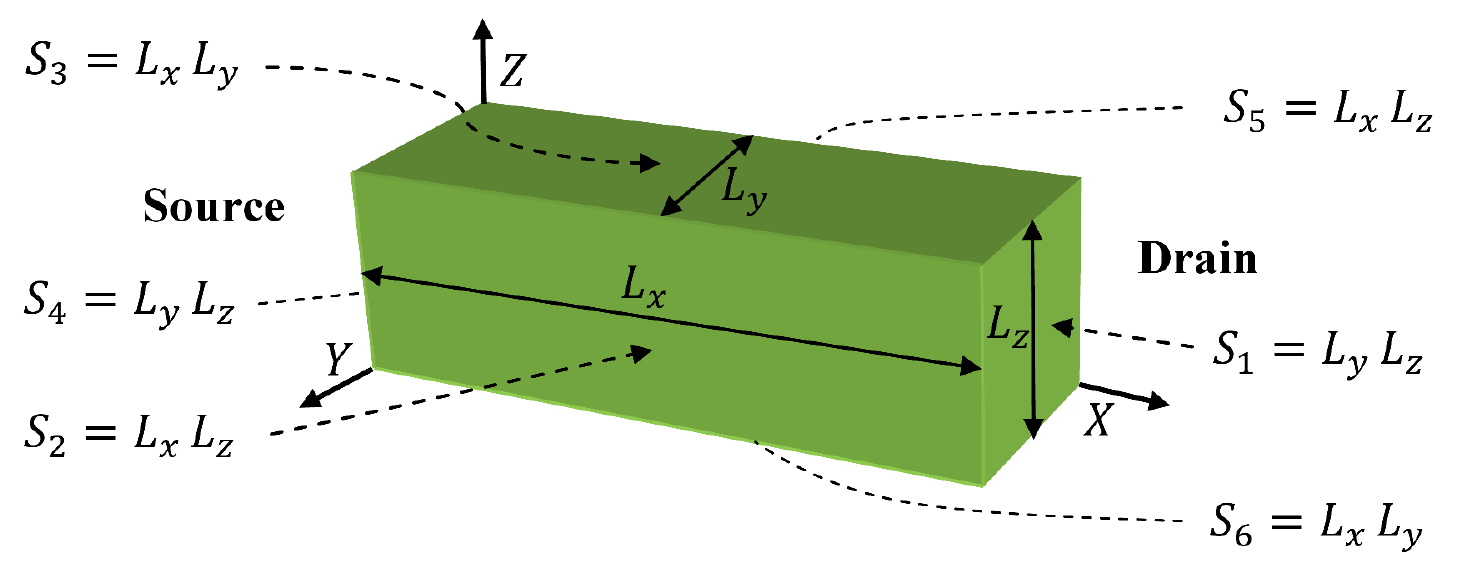}
\vspace*{8pt}
\caption{%
   \emph{Schematic representation of an arbitrary parallelepiped of volume $\Omega=L_{x}L_{y}L_{z}$ limited by the closed surface $S=\{S_1,...,S_6\}$. It can represent a two-terminal particle device where particle 
   transport takes place from source to drain.}}
\label{figure_1}
\end{figure}

Consider a parallelepiped of volume $\Omega=L_{x} L_{y} L_{z}$  limited by the surface $S$, which is composed of six rectangular surfaces $S=\{S_1,S_2,...,S_6\}$ (see Fig.~\ref{figure_1}). 
The only restriction on the surface $S$ is that it has to include the surface $S_i$ where the total current $I_i(t)$ wants to be computed. 
For each surface $S_i$, we define a scalar function $\Phi_i(\mathbf{r}')$ and its corresponding vector function $\mathbf{F}_i(\mathbf{r}')$, related through:
\begin {equation}
 \mathbf{F}_i(\mathbf{r}')=-\nabla \Phi_i(\mathbf{r}'),
 \label{equation_3}
\end {equation}
and fulfilling the equations:
\begin {equation}
 \nabla \cdot ( \epsilon (\mathbf{r}') \mathbf{F}_i(\mathbf{r}')) = -\nabla \cdot ( \epsilon (\mathbf{r}') \nabla \Phi_i(\mathbf{r}'))=0,
 \label{equation_4}
\end {equation}
with the following Dirichlet boundary conditions~\cite{coment9,coment10}:
\begin {equation}
 \Phi_i(\mathbf{r}')=1;\quad\mathbf{r}' \in S_i \;\; \textrm{   and   }  \; \; \Phi_i(\mathbf{r}')=0; \quad\mathbf{r}' \in S_{h\neq i}.
\label{equation_5}
\end {equation} 

The Ramo-Shockley-Pellegrini theorem then reads as follows: the total (displacement plus conduction) time-dependent electrical current measured on the surface $S_i$ can be written as: 
\begin{eqnarray}
 \langle I_i(t) \rangle = \langle \Gamma^{q}_i(t) \rangle + \langle \Gamma^{e}_i(t) \rangle,
\label{equation_25}
\end{eqnarray}
where the expressions for $\langle \Gamma^{q}_i(t) \rangle$ and $\langle \Gamma^{e}_i(t) \rangle$ read:
\begin{eqnarray}
 &&\langle \Gamma^{q}_i(t) \rangle = -\int_{\Omega } \mathbf{F}_i(\mathbf{r}') \cdot \overline{\mathbf{J}_c(\mathbf{r}',t)} dv',    \label{equation_25_q}\\
 &&\langle \Gamma^{e}_i(t) \rangle = \int_S \epsilon (\mathbf{r}')  \frac{d \overline{V(\mathbf{r}',t)}}{dt} \mathbf{F}_i(\mathbf{r}') \cdot d\mathbf{s}'. \label{equation_25_e}
\end{eqnarray}
In Eqs.~\eref{equation_25_q} and \eref{equation_25_e} we have defined the averaged scalar potential:
\begin {eqnarray}
 \overline{V(\mathbf{r}',t)} = \int_{\infty}dv |\Psi (\mathbf{r},t)|^2  U(\mathbf{r}',\mathbf{r},t),
\label{equation_15}
\end {eqnarray}
and the averaged current density:
\begin {eqnarray}
 \overline{\mathbf{J}_c(\mathbf{r}',t)} = \sum^{N}_{k=1}  \int_{\infty}d\bar{v}_k  \mathbf{J}_k(\mathbf{r},t)|_{\mathbf{r}_k=\mathbf{r}'},
\label{equation_23}
\end {eqnarray}
where $dv = dv_1 .. dv_{N}$ and $d\bar{v}_k = dv_1 .. dv_{k-1} dv_{k+1} .. dv_{N}$.
In Eq.~\eref{equation_15}, $U(\mathbf{r}',\mathbf{r},t)$ is the scalar potential that a test charge of unit charge at position $\mathbf{r}' = (x',y',z')$ would experience for a given probability distribution $|\Psi (\mathbf{r},t)|^2$.
In Eq.~\eref{equation_23}, 
\begin{equation}
 \mathbf{J}_k(\mathbf{r},t) = \frac {-iq_k}{2m_k}  \left( \psi^*(\mathbf{r},t){\nabla_k \psi(\mathbf{r},t)} - \psi(\mathbf{r},t){\nabla_k \psi^*(\mathbf{r},t)}\right)
 \label{current}
\end{equation} 
is the standard definition of the $k$-th component of the electrical current density.

The sum of the above contributions in Eqs.~\eref{equation_25_q} and \eref{equation_25_e} is equivalent to the sum of the standard conduction and displacement currents, i.e. 
$\langle I_i(t) \rangle = \langle \Gamma^{q}_i(t) \rangle + \langle \Gamma^{e}_i(t) \rangle = \langle \Upsilon^{c}_i(t) \rangle + \langle \Upsilon^{d}_i(t) \rangle$, where the last are defined as:
\begin{eqnarray}
 &&\langle \Upsilon^{c}_i(t) \rangle = \int_{S_i} \overline{\mathbf{J}_c(\mathbf{r}',t)} \cdot d\mathbf{s}', \label{equation_25_c}\\
 &&\langle \Upsilon^{d}_i(t) \rangle = \int_{S_i} \epsilon (\mathbf{r}')  \frac{d\overline{\mathbf{E}(\mathbf{r}',t)}}{dt} \cdot d\mathbf{s}',  \label{equation_25_d}
\end{eqnarray}
where  
\begin{equation}
 \overline{\mathbf{E}(\mathbf{r}',t)} = \int_{\infty}dv |\Psi (\mathbf{r},t)|^2  E(\mathbf{r}',\mathbf{r},t),
\end{equation}
and $E(\mathbf{r}',\mathbf{r},t)$ is the electrical field that a test charge of unit charge at position $\mathbf{r}'$ would experience for a given probability distribution $|\Psi (\mathbf{r},t)|^2$.
Please, notice that the terms $\Gamma^q_i(t)$ and $\Gamma^e_i (t)$ cannot be interpreted as the conduction, $\Upsilon^c_i (t)$, and displacement, $\Upsilon^d_i (t)$, currents, respectively. 
In particular, the term $\Gamma^q_i (t)$ includes not only the conduction current, but also part of the displacement current~\cite{QuantumRSP}.

\subsection{Modeling the Measuring Apparatus}
\label{section_2_2}
We want to characterize the interaction between an electronic device and a measuring apparatus designed to extract information on the electronic current that circulates across it. 
In this way, one can later model the effects (backaction) of a sequential measurement of the electrical current on the time evolution of the quantum system.

The standard approach to model the non-unitary evolution of the wavefunction $\psi(\mathbf{r},t)$ due to the backaction of the measuring apparatus consists in assuming that the wavefunction ``collapses'' into an eigenstates of the operator
that represents the observable being measured~\cite{vonNeuman,CohenTanudji}.
In fact, this kind of measurement, which is often referred to as von Neumann measurement, represents only a special class of all the possible measurements that can be made on quantum systems.
POVMs constitute an alternative (more general) class of operations specially suited, e.g., when some aspect of a quantum system wants to be continuously monitored.
These kind of measurements allow to provide information on the ensemble value of an observable without greatly disturbing the wavefunction of the system. 
The price to be paid is that very little information about the system is obtained in a single-shot measurement. 
In the following we derive a measuring protocol for the sequential measurement of the total electrical current based on POVM, here also referred to as weak-measurements~\cite{straightforward}.

Hereafter we focus only on the electrical current measured on surface $S_1$, and hence, for the sake of simplicity, we will omit from now on the subindex $i = 1$.
Let $\hat I(t)$ be the total current operator (whose mathematical form we do not know yet). 
This operator has to provide the same total current on surface $S_1$ that we computed from our quantum version of the RSP results, i.e.: 
\begin {eqnarray}
\label{oper1}
 \langle \hat I(t) \rangle = \left\langle\Gamma^{q}(t)\right\rangle + \left\langle\Gamma^{e}(t)\right\rangle = \int_{\infty}dv\int_{\infty}dv' \psi^*(\mathbf{r}',t) \bra{\mathbf{r}'} \hat I(t) \ket{\mathbf{r}} \psi(\mathbf{r},t).
\end {eqnarray}

Without the loss of generality, we select a volume $\Omega$ such that its lateral dimensions are much larger than the dimension along the transport direction. 
Then, the scalar functions $\Phi(\mathbf{r}')$ and its associated vector functions $\mathbf{F}(\mathbf{r}')$ defined in Eq.~\eref{equation_3} are the solutions of the potential and the electric field in a infinite parallel-plate capacitor, i.e.:
\begin {eqnarray}
\mathbf{F}(\mathbf{r}') = -\frac{1} {L_x} \mathbf{u}_x',   
\label{approx}
\end {eqnarray}
where the unit vector $\mathbf{u}_x'$ points in the positive $x$ (transport) direction of Fig.~\ref{figure_1}.
Since the surfaces $S_2,S_3,S_5$ and $S_6$ are very far from the active region, one can neglect their contribution to the current because the probability density there is roughly zero. 
Equation \eref{approx} is in fact equivalent to assume that the system of interest consists only of two terminals, i.e. the total current is negligible in all surfaces except on $S_1$ and $S_4$.

The component $\left\langle\Gamma^{e}(t)\right\rangle$ in Eq.~\eref{equation_25_e} can now be written as:
\begin{equation}
 \langle \Gamma^{e}_i(t) \rangle = -\frac{1} {L_x} \int_S \epsilon (\mathbf{r}')  \frac{d \overline{V(\mathbf{r}',t)}}{dt}  \mathbf{u}_x' \cdot d\mathbf{s}'.
\end{equation}
Since $\mathbf{u}_x' \cdot d\mathbf{s}'$ is non-zero only on surfaces $S_1$ and $S_4$, the aboe expression reduces to:
\begin{equation}
 \langle \Gamma^{e}_i(t) \rangle = \frac{1} {L_x} \int_{S_4} dy'dz' \epsilon (\mathbf{r}')  \frac{d \overline{V(\mathbf{r}',t)}}{dt} -  \frac{1} {L_x}\int_{S_1} dy'dz' \epsilon (\mathbf{r}')  \frac{d \overline{V(\mathbf{r}',t)}}{dt}.
\end{equation}
If we now choose the volume $\Omega$ such that surface $S$ (both $S=S_1$ or $S_4$) lies in the metallic contacts, the component $\left\langle\Gamma^{e}(t)\right\rangle$ in Eq.~\eref{equation_25_e} can be finally written as:
\begin{equation}
 \left\langle\Gamma^{e}(t)\right\rangle = -\frac{\epsilon}{L_x}  \frac{dV_{bias}(t)}{dt}, 
 \label{equation_25_e_3}
\end{equation}
where we have assumed that $\epsilon(0) = \epsilon(L_x)$ in the metallic regions does not depend on the transversal dimensions (y and z), 
and we defined the applied bias as: 
\begin{equation}
 V_{bias}(t) = \int_{S_1} dy'dz'  \overline{V(\mathbf{r}',t)} -  \int_{S_4} dy'dz' \overline{V(\mathbf{r}',t)}. 
\end{equation}

Next we focus on the term $\left\langle\Gamma^{q}(t)\right\rangle$ of Eq.~\eref{equation_25_q}, which taking into account Eq.~\eref{approx} can be written as:
\begin{eqnarray}
 \left\langle \Gamma^{q}(t)\right\rangle = \frac {1} {L_x} \int_{\Omega} dv' \; \overline{\mathbf{J}_c(\mathbf{r}',t)} \cdot \mathbf{u}_x' = 
   \sum^{N}_{k=1} \frac {1} {L_x} \int_{\infty} d\bar{v}_k  \int_{\Omega}dv_k  \mathbf{J}_k(\mathbf{r},t) \cdot \mathbf{u}_{x_k}. 
   \label{derivation_1}
\end{eqnarray}
Using the definition of $\mathbf{J}_k(\mathbf{r},t)$ in Eq.\eref{current}, the above equation can be written as:
\begin{equation}
 \left\langle \Gamma^{q}(t)\right\rangle = \sum^{N}_{k=1} \frac {1} {L_x} \int_{\infty} d\bar{v}_k  \int_{\Omega}dv_k \frac {-iq_k}{2m_k}  \left( \psi^*(\mathbf{r},t){\frac {\partial} {\partial x_k} \psi(\mathbf{r},t)} - \psi(\mathbf{r},t){\frac {\partial} {\partial x_k} \psi^*(\mathbf{r},t)}\right).
 \label{derivation_2}
\end{equation}
Equation \eref{derivation_2} can be then simplified by considering the following identity:
\begin {eqnarray}
 \psi(\mathbf{r},t)\frac {\partial} {\partial x_k} \psi^*(\mathbf{r},t) = \frac {\partial} {\partial x_k} |\psi(\mathbf{r},t)|^2 - \psi^*(\mathbf{r},t)\frac {\partial} {\partial x_k} \psi(\mathbf{r},t).
\end {eqnarray}
Introducing the above equality into Eq.~\eref{derivation_2} and taking into account the fact that the selected volume $\Omega$ is large enough so that its surface $S$ lies in an ideal metal, 
we realize that the total charge in that surface is ``instantaneously'' zero, i.e. the (dielectric relaxation) time needed to find local charge neutrality is roughly zero~\cite{MJavidMcGrawHillBook1963}.
Then, the following identity for the unidimensional charge density at surface $S$ holds:
\begin {eqnarray}
 \sum^{N}_{k=1} q_k \int_{\infty}d\bar{v}_k \int_{S_1} dy_k dz_k |\psi(\mathbf{r},t)|^2 = 0,
\label{charge_metal2}
\end {eqnarray}
and hence Eq.~\eref{derivation_2} finally reduces to:
\begin{equation}
 \left\langle \Gamma^{q}(t)\right\rangle = \sum^{N}_{k=1} \frac{1}{L_x} \int_{\infty}d\bar{v}_k \int_{\Omega}dv_k  \psi^*(\mathbf{r},t) \frac{-iq_k}{m_k}{\frac {\partial} {\partial x_k} \psi(\mathbf{r},t) }.
 \label{derivation_3}
\end{equation}

The comparison of Eqs.~\eref{equation_25_e_3} and \eref{derivation_3} with Eq.~\eref{oper1} already allows to write an expression for the total current operator $\bra{\mathbf{r}'} \hat I(t) \ket{\mathbf{r}}$ 
in the position representation:
\begin {eqnarray}
  {I}(\mathbf{r},\mathbf{r}',t)  =  \delta(\mathbf{r}'-\mathbf{r}) \left(  \frac {1} {L_x} \sum_{k=1}^{N}  \Theta^{x_{S4}}_{x_k}\Theta^{x_k}_{x_{S_1}} \frac{q_k}{m_k} p_{k_x}(\mathbf{r})
                                    - \frac{\epsilon}{L_x}  \frac{dV_{bias}(t)}{dt} \right),
\label{operator}
\end {eqnarray}
where $x_{S_1}$ and $x_{S_4}$ are the limits of the volume $\Omega$ in the $x$ (transport) direction (see Fig.~\ref{figure_1}), $\Theta^{x_{S4}}_{x_k} = \Theta(x_{S4} - x_k)$ and $\Theta^{x_k}_{x_{S_1}} = \Theta(x_k - x_{S_1})$ 
are Heaviside functions, and $p_{k_x}(\mathbf{r})$ is the longitudinal $k$-th component of the linear momentum operator in the position representation. 
Due to the presence of the Heaviside functions, the sum in Eq.~\eref{operator} effectively runs only over those particles, $N_S(t)$, whose associated reduced densities have a non-zero support in the volume $\Omega$.
The reader can be surprised that the current operator still depends on $L_x$. 
However, it is an artificial dependence compensated by the number of particles. As $L_x$ increases/decreases, the number of particles $N_S(t)$ is proportionally increased/decreased. 
Therefore, the total current remains independent of $L_x$ as required by expressions \eref{equation_25_q} and \eref{equation_25_e}. 

The eigenvalues $I(\mathbf{p}_x,t)$ of the current operator in Eq.~\eref{operator} depend on time and for many particles admit multiple combinations of the longitudinal components of the single-particle momentum eigenvalues: 
\begin{eqnarray}
 I(\mathbf{p}_x,t) = \frac{1}{L_x}  \sum_{k=1}^{N_S(t)} \frac{q_k}{m_k} p_{k_x},  
 \label{eigenvalues}
\end{eqnarray}
where $\mathbf{p}_x = \{p_{1_x},..,p_{N_{S_x}}\}$ collectively denotes the eigenvalues of the longitudinal components of the linear momentum operator.
Since the eigenstates of the current operator defined in Eq.~\eref{operator} can be always written as linear combinations of the total momentum eigenstates $\ket{\mathbf{p}} = \ket{\mathbf{p}_1} \otimes ... \otimes \ket{\mathbf{p}_{N}}$ 
(with $\ket{\mathbf{p}_k} = \ket{p_{k_x},p_{k_y},p_{k_z}}$), it will be useful to write the initial (before measurement) state at any time $t$ in terms of momentum eigenstates as~\cite{CohenTanudji}: 
\begin{eqnarray}
 \ket {\psi^o(t)} = \sum_{\mathbf{p}} g^o(\mathbf{p},t) \ket{\mathbf{p}},
\end{eqnarray}
with $g^o(\mathbf{p},t)=\braket{\mathbf{p}} {\psi^o(t)}$.

We are now in a position to propose a sequential weak-measurement protocol for the total electrical current. 
According to Ref.~\cite{povms}, the weak distortion introduced on the wavefunction by the measurement of the value $\mathcal{I}$ at time $t$ can be described by the generalized Gaussian operator $\hat W_\mathcal{I}$, i.e.:
\begin{equation}
 \hat W_{\mathcal{I}}= \frac {1}{C} \sum_{\mathbf{p}} e^{ -\frac {(I(\mathbf{p}_x,t) - \mathcal{I})^2} {2\sigma^2} } \hat M,
\label{weak}
\end{equation}
where $C$ is a normalization constant to guarantee $\sum_{j=-\infty}^{+\infty} \hat W_\mathcal{I} \hat W_\mathcal{I} = \hat{\mathbb{I}}$, and we have defined the projector $\hat M = \ket {\mathbf{p}} \bra {\mathbf{p}}$.
Instead of our measurement operators being projectors onto a single eigenstate of the total current, we choose them to be a weighted sum of projectors onto all current eigenstates, each one peaked about a different value of the observable $\mathcal{I}$.
In this respect, notice that degeneracy plays an important role in Eq.~\eref{weak} because a particular current value $\mathcal{I}$ can be reproduced through multiple momentum configurations.

The probability of finding $\mathcal{I}$ during a measurement at time $t$ is then:
\begin{equation}
 prob(\mathcal{I},t) = Tr(\hat W_{\mathcal{I}} \ket{\psi^o(t)} \bra{\psi^o(t)} \hat W_{\mathcal{I}}) =  \frac {1}{C^2}  \sum_{\mathbf{p}}   e^{ -\frac {(I(\mathbf{p}_x,t) - \mathcal{I})^2} {\sigma^2} }  |g^o(\mathbf{p},t)|^2,
\end{equation}
where $Tr\{\cdot\} = \sum_\mathbf{p}\langle \mathbf{p}|\cdot|\mathbf{p}\rangle$ denotes the trace over momentum space.  
We can now test if the average value of the current at one particular time $t$ computed with the weak-measurement formalism coincides with the value obtained with the projective (von Neumman) formalism. 
A simple calculation shows:
\begin{eqnarray}
 \langle \mathcal{I} \rangle = \sum_{\mathcal{I} = -\infty}^{+\infty}  \mathcal{I} \; prob(\mathcal{I},t)  = \sum_{\mathbf{p}}  I(\mathbf{p}_x,t)   |g^o(\mathbf{p},t)|^2 = \langle \hat I(t) \rangle,
\label{mean}
\end{eqnarray}
where we have used that $\frac{1}{C^2} \sum_\mathcal{I}  \mathcal{I}   e^{ -\frac{(I(\mathbf{p}_x,t) - \mathcal{I})^2} {\sigma^2} } = I(\mathbf{p}_x,t)$.
This result proves that the weak operator in Eq.~\eref{weak} provides the correct expectation value of Eq.~\eref{oper1} independently of the value of $\sigma$. 
However, Eq.~\eref{mean} was derived under the assumption that our electronic device was measured only once at (time $t$). 
For a multi-time measurement, the independence of $\langle \mathcal{I} \rangle$ on the value of $\sigma$ does not hold anymore because each measurement affects the evolution of the wavefunction $\ket{\psi(t)}$. 
In other words, we must combine the unitary evolution due to the time-dependent Schr\"odinger equation in Eq.~\eref{schrodinger} with the non-unitary evolution due to the sequential measurement process, which reads~\cite{povms}:
\begin{equation}
 i\frac{\partial}{\partial t} \ket {\psi^o(t)} = \frac{\ket {\psi^f(t)}} { \braket{\psi^f(t)}{\psi^f(t)} },
 \label{measured_state_n}
\end{equation}
where the after-mesurement (non-normalized) state, $\ket {\psi^f(t)}$, is: 
\begin{eqnarray}
 \ket {\psi^f(t)} =  \hat W_{\mathcal{I}} \ket {\psi^o(t)} =  \frac {1}{C} \sum_{\mathbf{p}}   e^{ -\frac {(I(\mathbf{p}_x,t) - \mathcal{I})^2} {2\sigma^2} }  g^o(\mathbf{p},t) \ket  {\mathbf{p}}.  
 \label{measured_state_un}
\end{eqnarray}

The weak-operator in Eq.~\eref{weak} has two interesting limits. The parameter $\sigma$ (in units of current) characterizes the ``strength'' of the measurement: $\sigma \rightarrow 0$ is close to a projective measurement of the total current, 
while $\sigma \rightarrow \infty$ means a measurement without distortion of the wavefunction.
For $\sigma \to \infty$ the Gaussian function in Eq.~\eref{weak} is much broader than $|g^o(\mathbf{p},t)|^2$ and hence this last can be considered a delta function centered at $\langle \hat I(t) \rangle$ 
(such that $\langle \mathcal{I} \rangle  =   \langle \hat I(t) \rangle$). 
Then, the probability of finding a particular value of the current $\mathcal{I}$ during a measurement is:
\begin{equation}
 prob(\mathcal{I},t) \approx  \frac {1}{C^2}   e^{ -\frac {(\langle \hat I(t) \rangle - \mathcal{I})^2} {\sigma^2} }  |g^o(\mathbf{p},t)|^2,
\end{equation}
and the state $\ket {\psi^o(t)}$ is not distorted during the measurement, i.e.:
\begin{equation}
 \ket {\psi^f(t)} \approx \ket {\psi^o(t)}.
\end{equation}
Contrarily, for $\sigma \to 0$, the Gaussian operator $\hat W_{\mathcal{I}}$ reduces to:
\begin{equation}
 \hat W_{\mathcal{I}}= \frac {1}{C} \sum_{\mathbf{p}} \delta(I(\mathbf{p}_x,t) - \mathcal{I}) \hat M,
\end{equation}
and hence the probability of finding the outcome $\mathcal{I}$ is:
\begin{equation}
 prob(\mathcal{I},t) \approx  \frac {1}{C^2}  \sum_{\mathbf{p}}  \delta(I(\mathbf{p}_x,t) - \mathcal{I})  |g^o(\mathbf{p},t)|^2.
\end{equation}
The after-measurement (non-normalized) state can be written as:
\begin{eqnarray}
 \ket {\psi^f(t)} \approx  \hat M_{\mathcal{I}} \ket {\psi^o(t)} =  \frac {1}{C} \sum_{\mathbf{p}}   \delta(I(\mathbf{p}_x,t)-\mathcal{I})   g^o(\mathbf{p},t) \ket  {\mathbf{p}}.
 \label{projective_f}
\end{eqnarray}
Notice again the important role of degeneracy in expression~\eref{projective_f}.

\section{Application to electronic devices}
\label{section_3}
\begin{figure}[th]
\centering
\includegraphics[width=\textwidth]{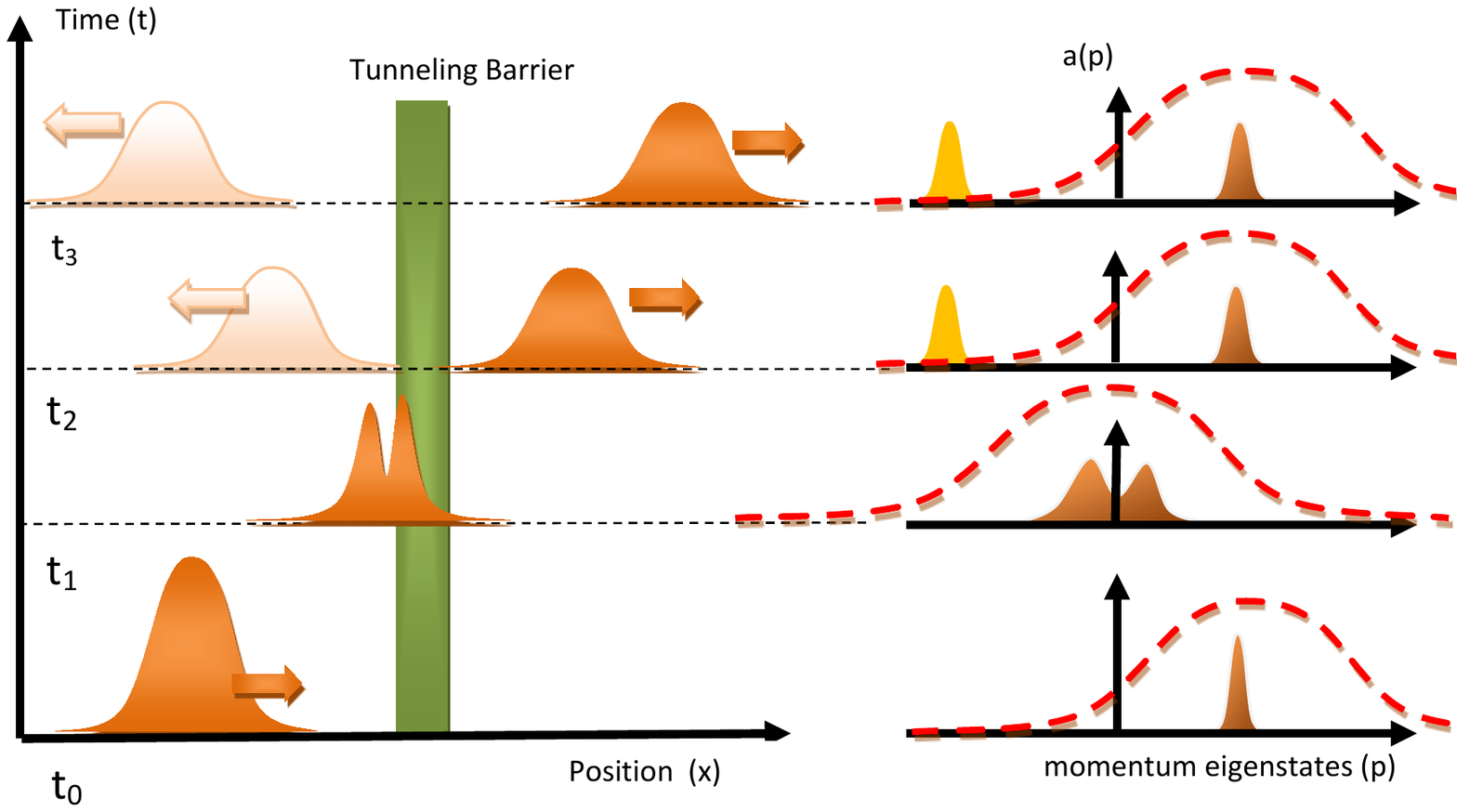}
\caption{%
    Schematic representation of the time-evolution of a (single-electron) wave-packet impinging upon a barrier. 
    (Left) The time-dependent wavepacket is spatially separated after time $t_2$ into transmitted and reflected parts. 
    (Right) Representation of the probability of the momentum eigenstates at different times. The dashed line correspond to the Gaussian function associated to the weak measurement in Eq.~\eref{weak}. 
    When the wavepacket is spatially separated, both in momentum and position space the backaction of the ammeter is reflected into the suppression of the reflected wavepacket. Hence after time $t=t_3$ 
    only positive electrical currents will be measured.}
\label{figure_2}
\end{figure}

Before carrying out any numerical calculation, let us discuss a very simple but elucidating example for the overall sequential measurement process. 
Consider an electron impinging onto a tunneling barrier. 
As depicted in Figure~\ref{figure_2}, we choose the value of $\sigma$ such that it is larger than the width of both the transmitted and reflected wavepackets (in the momentum space) at any time. 
As the incident wavepacket evolves in time, the momentum (and also spatial) support of the wavefunction splits into two non-overlapping components respectively associated to transmitted and reflected parts of the wavefunction. 
At time, $t=t_3$, the value of $\sigma$ becomes smaller than the width of the full (transmitted plus reflected) wavepacket. 
Therefore, as soon a given value of the current is measured, the wavepacket collapses into either its transmitted or reflected component.
In the example of figure~\ref{figure_2}, a positive value of the current is measured at time $t=t_3$, and hence, for later times $t>t_3$, the measured value of the current will be always positive.

The above example does not refer to a class of von Neumann measurements, but it is neither a strictly speaking continuous weak-measurement since the impact of the measuring process does sensibly affect the evolution of the quantum system. 
This example rather emphasizes the crucial role of the Gaussian width $\sigma$ in Eq.~\eref{weak}, which encompasses the transition between continuous and single-shot projective measurements. 
Even far from the projective regime, the dynamics of quantum systems can be greatly affected by the action of weak operators, and hence it is of paramount importance to properly characterize the measuring apparatus. 
This will be clearer from the following numerical example.

\subsection{A resonant tunneling device under continuous monitoring of the electrical current}
\label{section_3_2}
We finally put in practice the measuring protocol derived in the previous subsection in combination with the electron transport simulator BITLLES~\cite{BITLLES_1}.
The BITLLES simulator allows to perform both classical~\cite{Cresults_1} and quantum~\cite{Qresults_1} simulations of electronic devices, 
respectively using a classical Monte Carlo technique to solve the Boltzmann transport equation~\cite{MonteCarlo_C},
and a quantum Monte Carlo algorithm to solve the many-body Schr\"odinger equation based on the use of conditional wavefunctions~\cite{conditional_1}.
In both cases the self-consistency of the electronic equations of motion and the many-body Poisson equation is fulfilled~\cite{selfconsistent_1} in combination with a proper set of 
time-dependent boundary conditions~\cite{boundary_1} and injection algorithms~\cite{injection}.

We have recently implemented Eq.~\eref{weak}, at the single-particle level, into the BITLLES simulator using regular fast Fourier transforms.
For a single electron, i.e. $N=1$, the eigenvalues of the current operator in Eq.~\eref{operator} coincide (up to a constant) with those of the $x$-component of the momentum operator:
\begin{eqnarray}
 I(p_x) = -\frac{p_x}{L_x} .  
 \label{single_particle}
\end{eqnarray}
Expression \eref{single_particle} establishes now a unequivocal relation between the total electrical current and the linear momentum eigenstates. 
Notice that eigenstates of the current operator are now also single-particle momentum eigenstates: $\ket{\mathbf{p}} = \ket{p_{x},p_{y},p_{z}}$, and hence
a given outcome of the (measured) current, $\mathcal{I}$, is associated (one-to-one) to a particular momentum eigenstate such that $I(p_x) = \mathcal{I}$.
Therefore, in the single-particle limit, the numerical implementation of Eqs.~\eref{measured_state_n} and \eref{measured_state_un} is greatly simplified,
and for a given value of $\sigma$, the computational cost (if any) of the measuring protocol is directly associated with the manipulation of direct and inverse (fast) Fourier transforms. 

As a very first example, in this work we consider a single-particle resonant tunneling diode in one dimension. 
The double barrier structure has a $0.4$nm well and barrier heights and widths of $0.5$eV and $0.4$nm respectively. 
An electron impinging into the nanostructure is initially represented by a Gaussian wavepacket with positive linear momentum corresponding to an average energy of $0.25$eV such that it is in resonance with the double barrier.  
We consider the initial spatial dispersion of the electronic wavepacket to be $30$nm. 
\begin{figure}[th]
\centering
\includegraphics[width=\textwidth]{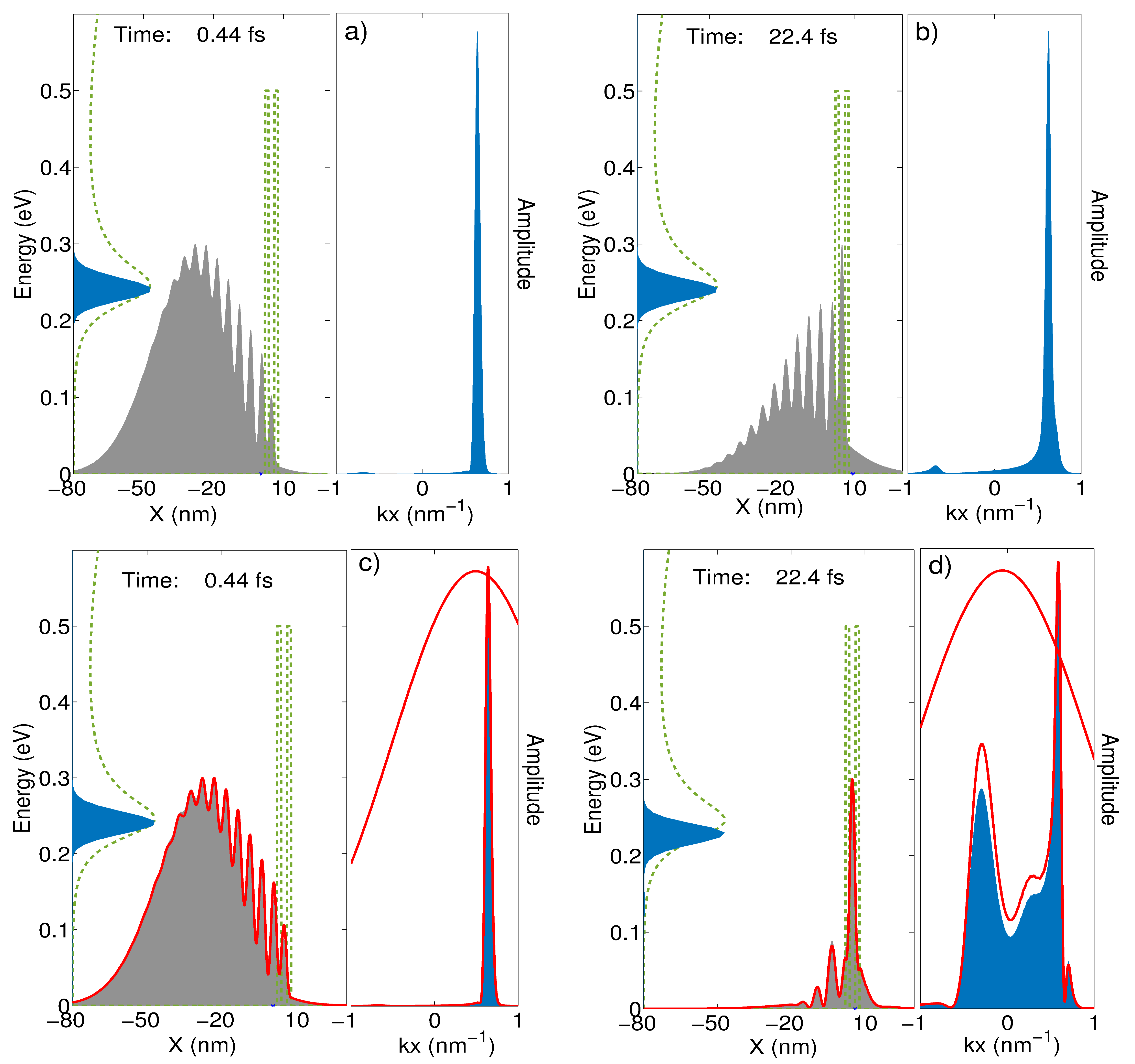}
\caption{%
   In (a) and (b): time-evolution of a (single-electron) wavepacket impinging upon a double barrier structure. (Left panels) Wavepacket in position representation (in gray), 
   double barrier structure and transmission coefficient as a function of energy (in dotted green lines), and energy distribution of the wavepacket in blue. (Right panels) Wavepacket in momentum representation.
   In (c) and (d): the same information as in (a) and (b) but now for the total electrical current being monitored with the POVM of Eq.~\eref{weak} with parameters $\sigma = 2\cdot 10^9m^{-1}$ and $\tau = 4\cdot 10^{-16}s$.
   Continuous red lines refer respectively to the after-measurement wavepacket in position, $\psi^f(x,t)$, and momentum, $g^f(p,t)$, representations.}
\label{figure_3}
\end{figure}

In Fig.~\ref{figure_3}, we show snapshots, at times (a) $t=0.44$fs and (b) $t=22.4$fs, of the evolution (without measurement) of the wavepacket $\psi^o(x,t)$ (in gray) along with its momentum (and energy) distributions $g^o(p,t)$ in blue. 
In panels (c) and (d), the same information is plot for the case in which the current generated by the wavepacket is being monitored with an ammeter with the following parameters: $\sigma = 2\cdot 10^9m^{-1}$ and $\tau = 4\cdot 10^{-16}s$.
The after-measurement wavepacket (both in position, $\psi^f(x,t)$, and momentum, $g^f(p,t)$, spaces) together with the Gaussian distribution $e^{ -\frac {(I(\mathbf{p}_x) - \mathcal{I})^2} {2\sigma^2} }$ are plot in red.

As can be seen, at the initial time (see Figs.~\ref{figure_3}.a and \ref{figure_3}.c) there is no appreciable differences between measured and non-measured scenarios. 
Indeed, for the given value of $\sigma$, the effect of the ammeter on the time-dependent wavepacket is rather innocuous ``instantaneously'' (at any time). 
It is, however, the insistent (almost continuous) weak-action of the ammeter on the quantum system what generates important differences at later times.
Notice, for instance, the effects of the ammeter on the wavepacket by comparing Figs.~\ref{figure_3}.b and \ref{figure_3}.d at time $t=22.4fs$. 
The instantaneous (initially negligible) effect of the measurement process on electron dynamics turns into a strong perturbation after being applied hundreds of times in a few femtoseconds.

\begin{figure}[th]
\centering
\includegraphics[width=\textwidth]{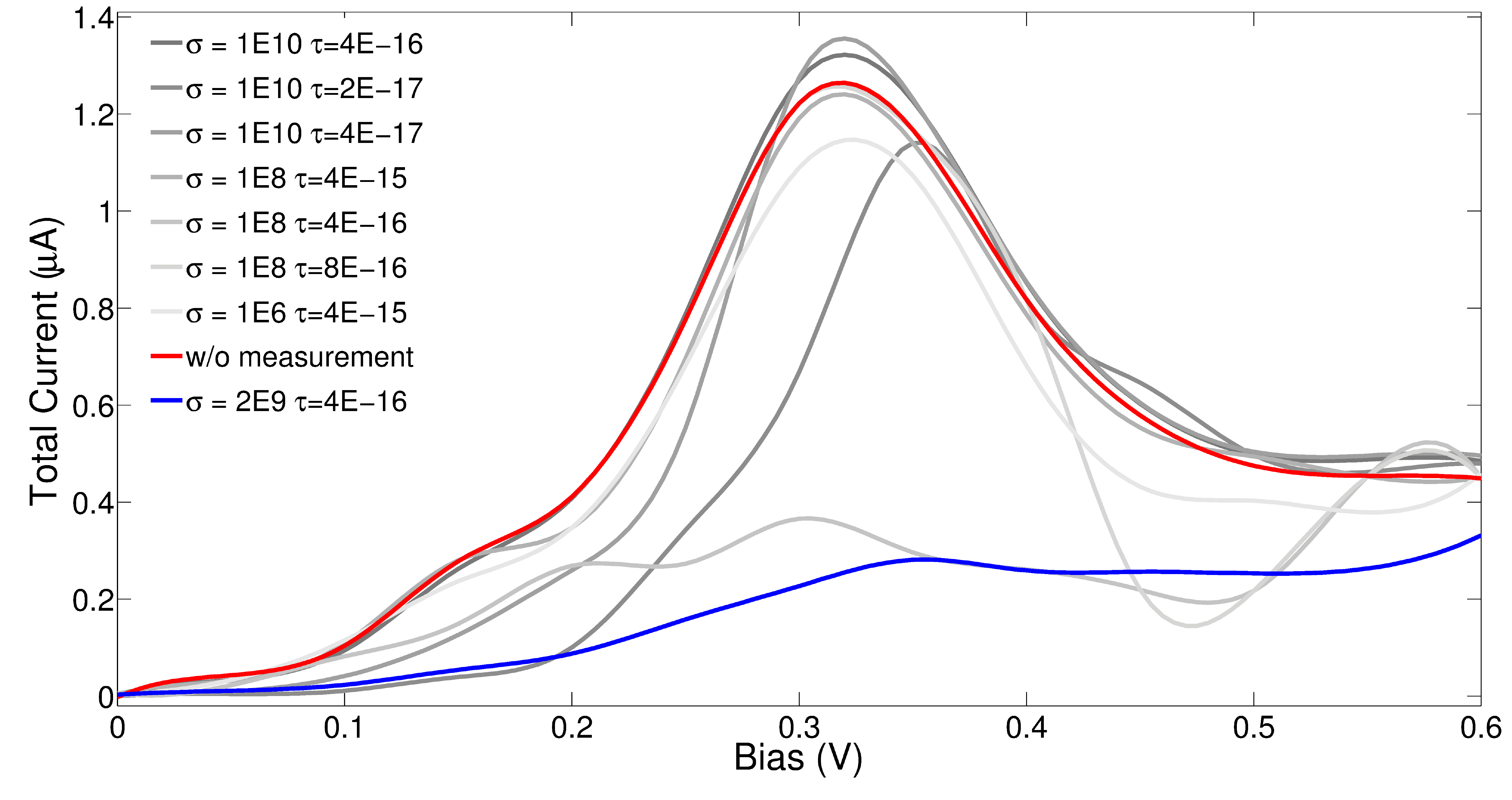}
\caption{%
   Current-voltage characteristics for a resonant tunneling diode under sequential measurement of the total electrical current for different configurations of the parameters $\sigma$ and $\tau$.
   In particular, in blue and red are respectively the cases for $\sigma = 2\cdot 10^9m^{-1}$ and $\tau = 4\cdot 10^{-16}s$ and without measurement, for which the dynamics has been depicted in Fig.~\ref{figure_3}.
   }
\label{figure_5}
\end{figure}

Further, as seen in Fig.~\ref{figure_5}, the effects of ``continuously'' monitoring the electrical current on the dynamics of electrons has a direct impact on the current-voltage characteristic of the resonant tunneling diode. 
Notice in particular the suppression of resonance peak and hence also of the negative conductance region for the case described above (i.e. $\sigma = 2\cdot 10^9m^{-1}$ and $\tau = 4\cdot 10^{-16}s$). 
The sensibility of particle dynamics to the values of $\sigma$ and $\tau$ is rather dramatic, and very different current-voltage characteristics can be found by slightly modifying these parameters.
Hence, the importance of properly characterizing (parameterizing) the measuring apparatus, even in the regime considered of weak disturbance.  

\section{Conclusions}
\label{section_4}
In this work we have derived a theoretical sequential weak-measurement protocol for the total, conduction plus displacement, electrical current.
Starting from an extended version of the Ramo-Schockley-Pellegrini theorem for quantum systems, we have been able to write an expression for the total current operator (see Eq.~\eref{operator}). 
The eigenvalues and vectors of this operator are in general degenerate and coincident (up to a constant) with the linear momentum eigenstates (see expression \eref{eigenvalues}).
Based on this finding, it is then simple to mathematically construct a generalized POVM which allows to formulate a sequential weak-measurement protocol for the total electrical current (see Eq.~\eref{weak}).
This POVM allows to provide information on the ensemble value of the total current in a range of situations depending on the parameters $\sigma$ and $\tau$. 
The Gaussian width $\sigma$ controls the ``strength'' of the measurement and $\tau$ defines the lapse of time between measurements.
The combination of these two parameters allows to go from a continuous weak-measurement scheme to a von Neuman projective measurement. 
The (either weak or strong) distortion on the quantum system introduced as a particular value of the total current $\mathcal{I}$ is measured is described by the operator $\hat W_\mathcal{I}$ through Eq.~\eref{measured_state_n}.

We have implemented the resulting measuring protocol for the total current at the single-particle level into the particle transport simulator BITLLES.
Specifically, equation \eref{weak} together with values for the measurement strength $\sigma$ and the time interval between measurements $\tau$, defines a computational algorithm for the interplay between unitary and non-unitary 
evolutions in systems being ``continuously'' monitored.
Numerical results for a resonant tunneling diode show that the effects of the measuring apparatus (even for large values of $\sigma$) have an important impact on single-electron dynamics if the measurement scheme is close to be continuous 
(i.e. for small values of $\tau$). 
Furthermore, we noticed a great sensibility of the current-voltage characteristics of the resonant tunneling diode to slight variations of the parameters $\sigma$ and $\tau$.
Hence, the importance of properly characterizing the measuring apparatus, even if it is designed to actuate on the weak-perturbation regime.  
In this respect, alternative modeling of such type of \emph{continuous} measurement for the displacement and particle current in electronic devices without POVMs has been recently presented~\cite{PRL_measure}.
There, the authors provide an explicit simulation of the interaction between the electrons in the quantum system and those in the measuring apparatus.

\section*{Acknowledgments}
The authors would like to thank Prof. Xavier Oriols for useful discussions on the quantum measurement theory. 
G. A. acknowledges financial support from the Beatriu de Pin\'os program through the Project: 2014 BP-B 00244. 
F. L. T. acknowledges support from the DOE under grant DE-FG02-05ER46204.


\begin{thebibliography}{10}
\bibitem{coment9}
One can argue that the functions $\mathbf{F}_i(\mathbf{r}')$ do not fulfill expression \eref{equation_3} in the intersection between the surface $S_i$ (with $\Phi_i(\mathbf{r}')=1$) and rest of surfaces $S_{h\neq i}$ (with $\Phi_i(\mathbf{r}')=0$). 
However, one can always redefine the functions $\Phi_i(\mathbf{r}')$ and $\mathbf{F}_i(\mathbf{r}')$ on a new volume $\Omega^*$  equal to the previous $\Omega$ except for the ``offending'' points. 
See for example p. 41 in Ref.~\cite{Jakson} for a similar discussion.
\bibitem{coment10}
It is argued in the literature that $\Phi_i(\mathbf{r}')$  and $\mathbf{F}_i(\mathbf{r}')$ are the scalar potential and the electric field, respectively, when there is no particles in the volume $\Omega$. 
However, we will avoid this definition because it can lead to misleading conclusions. Notice for instance that $\Phi_i(\mathbf{r}')$ has no units and $\mathbf{F}_i(\mathbf{r})$ is the inverse of a distance.
\end{thebibliography}
\end{document}